\documentclass[10pt,twocolumn,twoside] {IEEEtran}

\usepackage{cite}

\ifCLASSINFOpdf

\else
   \usepackage[dvips]{graphicx}
   \DeclareGraphicsExtensions{.eps}
\fi
\usepackage[cmex10]{amsmath}
\usepackage{algorithmic}
\usepackage{array}
\usepackage{amsthm}
\theoremstyle{remark}

\usepackage{booktabs}
\usepackage{bm}
\usepackage{tabularx}
\begin{document}
\title{ A Simplified Sub-Nyquist Receiver Architecture for Joint DOA and Frequency Estimation}
\author{Liang~Liu
        and~Ping~Wei
\thanks{The authors are with the Center for Cyber Security, School of Electronic Engineering, University of
Electronic Science and Technology of China, Chengdu 611731, China (e-mail: liu\_yinliang@outlook.com; pwei@uestc.edu.cn).}}

%\markboth{Journal of \LaTeX\ Class Files,~Vol.~11, No.~4, December~2012}%
%{Shell \MakeLowercase{\textit{et al.}}: Bare Demo of IEEEtran.cls for Journals}

\maketitle

\begin{abstract}
Joint estimation of carrier frequency and direction of arrival (DOA) for multiple signals has been found in many practical applications such as Cognitive Radio (CR). However, Nyquist sampling mechanism is costly or implemented due to wide spectrum range. Taking advantage of sub-Nyquist sampling technology, some array receiver architectures are proposed to realize joint estimation of carrier frequency and DOA. To further decrease equivalent sampling rate and hardware complexity, we propose a simplifying receiver architecture based on our previous work. We come up with joint DOA and frequency estimation algorithms for the novel architecture. The simulations demonstrate that the receiver architecture and the proposed approaches are feasible.
\end{abstract}


\IEEEpeerreviewmaketitle

\section{Introduction}
\IEEEPARstart{N}{owadays}, both carrier frequency and direction of arrival (DOA) are needed in some applications, such as Cognitive Radio (CR) aiming at solving the spectral congestion \cite{Haykin2005, Yucek2009, Mishali2011a, Sun2013, Cohen2014}. The most important function of CRs is to autonomously exploit locally unused spectrum to provide new paths to spectrum access. Therefore, spectrum sensing is an essential part of CRs. The conventional spectrum opportunity only contains three dimensions of the spectrum space: frequency, time, and space. However, with the advancement in array processing technologies \cite{Krim1996, Schmidt1986, Roy1986}, the new dimension, DOA, also creates new spectrum opportunities. Joint frequency spectrum and spatial spectrum would enhance the performance of CRs.

Recently, significant effort have been made towards jointly estimation of carrier frequencies and their DOAs \cite{Lemma1998, Lemma2003}. An obvious drawback is that they require additional pairing between the carrier frequencies and the DOAs. Besides, both works assume that the signal is sampled at least at its Nyquist rate. The main challenge of CRs lies in wideband signal procesing for their costly or even unreachable Nyquist rate sampling. The distribution range of the spectrum under monitoring  is from 300 MHz to several GHz \cite{Haykin2005, Yucek2009, Mishali2011a, Sun2013, Cohen2014}. It leads to high Nyquist sampling rate and a large number of sampling data to process.

Fortunately, sub-Nyquist sampling technology can reconstruct a multiband signal from its sub-Nyquist samples \cite{Mishali2011,Mishali2010,Eldar2009,Mishali2009}. Latterly, some joint DOA and carrier frequency estimation methods are proposed at sub-Nyquist sampling rates. \cite{Ariananda2013} proposes a structure, i.e. a linear array by employing a multi-coset sampling at the output of every sensor. This method compresses the wide-sense stationary signal in both the time domain and spatial domain. To simplify the hardware complexity, \cite{Kumar2014} uses an additional identical delayed channel at the output of every sensor. But there are ambiguities  during pairing with their corresponding DOAs in an underlying uniform linear array (ULA) scenario. To solve the pairing issue, \cite{ Kumar2015} proposes a structure with the hardware complexity identical to that of \cite{Kumar2014}. However, those papers do not give a unified signal reception model. \cite{Stein2015} presents two joint DOA and carrier frequency recovery approaches for an L-shaped ULA scenario. In \cite{Liu2016}, we propose a new array receiver architecture associated with two sub-Nyquist sampling based methods for simultaneously estimate the frequencies and DOAs of multiple narrowband far-field signals impinging on a ULA, where signals’ carrier frequencies spread around the whole wide spectrum. The architecture is complex due to every sensor following a multi-channel sub-Nyquist sampling receiver.

We consider a scene as \cite{Liu2016} in this paper.  For reducing the complexity of receiver, we propose a simplified array receiver architecture. For this model, we propose a unified formula and methods for joint estimation of DOA and carrier frequency.

%This paper is organized as follows: in Section II, we review the data model and main conclusions of array signal model with sub-nyquist. In Section III, we present our proposed simplified receiver architecture form and propose methods for joint DOA and frequency estimation. Section IV carries out the simulation experiment and finally we give the conclusions of this paper in Section V.

The following notations are used in the paper.  ${\left(  \cdot  \right)^{\rm T}}$ and ${\left(  \cdot  \right)^{\rm H}}$  denote the transpose and Hermitian transpose, respectively. $E\left(  \cdot  \right)$ stands for the expectation operator.  ${x_j}$ is the  $j$th entry of a vector ${\bf{x}}$.  ${{\bf{A}}_i}$ and ${A_{ij}}$ are the $i$th column and $(i,j)$th entry of a matrix ${\bf{A}}$, respectively. $ \otimes$  denotes the Hadamard product.  ${{\bf{I}}_M}$ stands for an $M \times M$ identity matrix.

\section{Array Signal Model with Sub-Nyquist Sampling}

In \cite{Liu2016}, we proposed an array signal receiver architecture and the corresponding signal reception model, which introduces sub-Nyquist sampling technology. In this letter, on one hand, the proposed architecture is the simplified form of the previous architecture, on the other hand, we will take advantage of the previous model when estimation algorithm deducing. Therefore, we review the main conclusions of \cite{Liu2016} in this section.

Consider $K$ narrowband far-field signals impinging on a ULA composed of $M$ $(M>K)$ sensors. Our previous receiver architecture applies  multi-coset sampling \cite{Mishali2009}. And every array sensor is followed by same $P$ delay branches. All the ADCs are synchronized and samples at a sub-Nyquist sampling rate of ${f_s} = {{{f_N}} \mathord{\left/
 {\vphantom {{{f_N}} L}} \right.
 \kern-\nulldelimiterspace} L}$, where ${f_N} = {1 \mathord{\left/
 {\vphantom {1 {{T_N}}}} \right.
 \kern-\nulldelimiterspace} {{T_N}}}$ is the Nyquist sampling rate. The  constant set  $C=[c_1,c_2,\cdots,c_P]$ is the sampling pattern where
$0 \le {c_1} < {c_2} <  \cdots  < {c_P} \le L - 1$. ${y_{mp}}\left[ n \right]$ denotes the sampled signal corresponding to the $m$th sensor, $p$th branch.
The matrix output of all branches of all sensors is given by
\begin{align}\label{Yf}
{\bf{Y}}\left( f \right) &= \left( {{\bf{A}} \otimes {\bf{B}}} \right)\overline {\bf{S}} \left( f \right)  + \left( {{{\bf{I}}_M} \otimes {\bf{B}}} \right)\widehat {\bf{N}}\left( f \right)\\
 &\buildrel \Delta \over= {\bf{G}}\overline {\bf{S}} \left( f \right)+{{\bf{I}}_{\bf{B}}}\widehat {\bf{N}}\left( f \right), f \in \mathcal{F} \buildrel \Delta \over =  \left[ {0,\frac{1}{{LT}}} \right),\label{Yf2}
\end{align}
where ${B_{il}} = \frac{1}{{\sqrt L }}\exp \left( {j\frac{{2\pi }}{L}{c_i}l} \right)$, ${{{A}}_{mk}} = \exp \left( { - j{\phi _k}\left( {m - 1} \right)} \right)$ is the $mk$th element of the steer array ${\bf{A}}$, where spatial phase
\begin{align}\label{Phi}
{\phi _k} = \frac{{2\pi d\sin \left( {{\theta _k}} \right)}}{{{c \mathord{\left/
 {\vphantom {c {{f_k}}}} \right.
 \kern-\nulldelimiterspace} {{f_k}}}}},
\end{align}
where ${\theta _k}$ and ${f_k}$ are the DOA and the center frequency of ${s_k}\left( t \right)$, respectively.
$\overline {\bf{S}} \left( f \right) = {\left[ {\overline {\bf{S}} _1^{\rm{T}}\left( f \right),\overline {\bf{S}} _2^{\rm{T}}\left( f \right), \cdots ,\overline {\bf{S}} _K^{\rm{T}}\left( f \right)} \right]^{\rm{T}}}$, ${\overline {\bf{S}} _k}\left( f \right) = {\left[ {{S_{k1}}\left( f \right),{S_{k2}}\left( f \right), \cdots ,{S_{kL}}\left( f \right)} \right]^{\rm{T}}}$, ${S_{kl}}\left( f \right) = {S_k}\left( {f + \frac{{l - 1}}{{LT}}} \right)$, ${S_k}\left( f \right)$ is the Fourier transform of ${s_k}\left( t \right)$. ${\bf{s}}\left( t \right){\rm{ = }}{\left[ {{s_1}\left( t \right),{s_2}\left( t \right), \cdots ,{s_K}\left( t \right)} \right]^{\rm{T}}}$ is the vector of all signal values.
Because ${s_k}\left( t \right)$ is a  narrowband signal, there is one, and only one frequency band which is occupied in ${\overline{{\bf{S}}}_k}\left( f \right)$. Further, ${\overline{{\bf{S}}}_{k}}\left( f \right)$ is a sparse vector of length $L$ when $k$ is fixed and there is one, and only one index ( marked as ${l_k}$), which is activated.
${\bf{Y}}\left( f \right) = {\left[ {{\bf{Y}}_1^{\rm{T}}\left( f \right),{\bf{Y}}_2^{\rm{T}}\left( f \right), \cdots ,{\bf{Y}}_M^{\rm{T}}\left( f \right)} \right]^{\rm{T}}}$. The $p$th element of ${Y_{mp}}\left( f \right) = \sqrt L T_N{Y_{mp}}\left( {{e^{j2\pi fT}}} \right)$, which is the discrete-time Fourier transform  of the signal ${y_{mp}}\left[ n \right]$ except a coefficient difference $\sqrt L T_N$.
$\widehat {\bf{N}}\left( f \right) = {\left[ {\widehat {\bf{N}}_1^{\rm{T}}\left( f \right), \cdots ,\widehat {\bf{N}}_M^{\rm{T}}\left( f \right)} \right]^{\rm{T}}}$, ${\widehat {\bf{N}} }_m\left( f \right) = \left[ {{N_{m1}}\left( f \right), \cdots ,{N_{ml}}\left( f \right)} \right]^{\rm T}$, ${N_{kl}}\left( f \right) = {N_k}{f + \frac{{l - 1}}{{LT}}}$ is the Fourier transform of ${n_k}\left( t \right)$. ${\bf{n}}\left( t \right)= {\left[ {{n_1}\left( t \right), \cdots ,{n_M}\left( t \right)} \right]^{\rm{T}}} $ is the noise vector, which subjects to the zero-mean circular complex Gaussian distribution with covariance matrix  ${\sigma}^2 \bf{I}_M$.

\section{Proposed receiver architecture and joint DOA and frequency estimation algorithm}

\subsection{Proposed receiver architecture}
\begin{figure}[!t]
\centering
\includegraphics[width=2.0in]{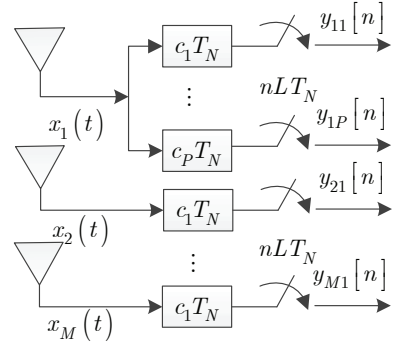}
\caption{Proposed receiver architecture.}
\label{figArcPart}
\end{figure}

To largely decrease hardware complexity, we design the simplified receiver architecture  when achieving joint frequency and DOA estimate. This architecture is set up based on the previous architecture. The main difference between the two architectures is that the former only reserves all branches of one array sensor and whole same branch of all sensors. The proposed receiver architecture is shown in Fig.\ref{figArcPart}. Without loss of generality, we select all branches of the first sensor and whole first branch of all sensors in the Fig.\ref{figArcPart}. Namely, our output is ${\bf{W}}\left( f \right) = {\left[ {{Y_{11}}\left( f \right),{Y_{12}}\left( f \right), \cdots ,{Y_{1P}}\left( f \right),{Y_{21}}\left( f \right), \cdots ,{Y_{M1}}\left( f \right)} \right]^{\rm T}}$. We define a $(M + P - 1)\; \times MP$ matrix ${\bf{J}}$, where ${J_{ij}} = 1$ for $\;i = 1, \cdots ,P$, and $ j=i$; or $i = P + 1, \cdots ,M + P - 1$, and $j={1 + iP - {P^2}}$; else ${J_{i,j}} = 0$ for else.
We have ${\bf{W}}\left( f \right) = {\bf{JY}}\left( f \right)$. According to (\ref{Yf}), we have
\begin{align}\label{Ypf}
{\bf{W}}\left( f \right) = {\bf{H}}\overline {\bf{S}} \left( f \right) + {\bf{J}}{{\bf{I}}_{\bf{B}}}\widehat {\bf{N}}\left( f \right),f \in\mathcal{F},
\end{align}
where ${\bf{H}} = {\bf{J}}\left( {{\bf{A}} \otimes {\bf{B}}} \right) = {\bf{JG}}$. Combing ${{\bf{I}}_{MP}} = {{\bf{I}}_{\bf{B}}}{\bf{I}}_{\bf{B}}^{\rm H}$ in \cite{Liu2016}, we have
\begin{align}\label{Noise}
{\bf{J}}{{\bf{I}}_{\bf{B}}}\widehat {\bf{N}}\left( f \right){\left( {{\bf{J}}{{\bf{I}}_{\bf{B}}}\widehat {\bf{N}}\left( f \right)} \right)^{\rm H}}{\rm{ = }}{\sigma ^2}{{\bf{I}}_{M + P - 1}}.
\end{align}

\subsection{Algorithm  Based on Individual Estimate}
\subsubsection{ Spatial Phase Estimate}

We denote the outputs of the 1st branch of all sensors as ${ {\bf{Q}} }\left( f \right) = {\left[ {\begin{array}{*{20}{c}}
{{{ {\bf{Y}} }_{11}}\left( f \right)}& \cdots &{{{ {\bf{Y}} }_{M1}}\left( f \right)}
\end{array}} \right]^{\rm T}}$. According to \cite{Liu2016}, we have the following equation.
\begin{align}\label{ASub}
{{\bf{Q}}}\left( f \right) = {\bf{A}}{{\bf{Z}}}\left( f \right) + \widehat {\bf{N}}_1\left( f \right),
\end{align}
where ${{\bf{Z}}}\left( f \right) = {\left[ {\begin{array}{*{20}{c}}
{\sum\limits_{l = 1}^L {{{{B}}_{1l}}{S_{1l}}\left( f \right)} }& \cdots &{\sum\limits_{l = 1}^L {{{{B}}_{1l}}{S_{Kl}}\left( f \right)} }
\end{array}} \right]^{\rm{T}}}$.
Because ${{\bf{S}}_{k}}\left( f \right)$ is a $1$-sparse vector of length $L$, and the  activated index is $l_k$. We can simplify ${{\bf{Z}}}\left( f \right)$ as ${\bf{Z}}\left( f \right) = {\left[ {\begin{array}{*{20}{c}}
{{B_{1{l_1}}}{S_{1{l_1}}}\left( f \right)}& \cdots &{{B_{1{l_K}}}{S_{K{l_K}}}\left( f \right)}
\end{array}} \right]^{\rm{T}}}$. (\ref{ASub}) is a standard array reception model, there are many existing method to get $\phi$, such as MUSIC, ESPRIT, and so on. Further, we can get the least square solution of ${ {{ {\bf{Z}} }}  }\left( f \right)$,
\begin{align}\label{ALS}
{ {{ {\bf{Z}} }}  }\left( f \right) = {\bf{A}} ^\dag { {\bf{Q}} }\left( f \right).
\end{align}

\subsubsection{ Frequency Estimate}
According to \cite{Liu2016} section III part B, the output of all  branches of  the 1st sensor is
\begin{align}\label{BSub}
{{\bf{Y}}_1}\left( f \right) = {\bf{B}}{\overline {\bf{X}} _1}\left( f \right),
\end{align}
where ${\overline {\bf{X}} _1}\left( f \right) = {\left[ {\begin{array}{*{20}{c}}
{\sum\limits_{k = 1}^K {{{{A}}_{1k}}{S_{k1}}\left( f \right)} }& \cdots &{\sum\limits_{k = 1}^K {{{{A}}_{1k}}{S_{kL}}\left( f \right)} }
\end{array}} \right]^{\rm{T}}}$. Since ${{\bf{S}}_{k}}\left( f \right)$ is a $1$-sparse vector of length $L$, ${\overline {\bf{X}} _1}\left( f \right)$ is $K$-sparse vector of length $L$. We denote the support set of ${\overline {\bf{X}} _1}\left( f \right)$ as $\Omega$. We can use the CTF algorithm to solve (\ref{BSub}) to obtain $\Omega $. Then, we hold
\begin{align}
{\overline {\bf{Y}} _1}\left( f \right) = {\bf{B}}{\overline {\bf{X}} _1}\left( f \right) = {{\bf{B}}_\Omega }{\left( {{{\overline {\bf{X}} }_1}} \right)_\Omega }\left( f \right).
\end{align}
Further, we can get the least square solution of ${\left( {{\overline {\bf{X}} _1}} \right)_\Omega }\left( f \right)$,
\begin{align}\label{BLS}
{\left( {{\overline {\bf{X}} _1}} \right)_\Omega }\left( f \right) = {\bf{B}}_\Omega ^\dag {\overline {\bf{Y}} _1}\left( f \right).
\end{align}
%Then, we can obtain ${\overline{f} }$.

%If we define ${\bf{P}}\left( f \right) = {\left[ {\begin{array}{*{20}{c}}
%{{Y_{11}}\left( f \right)}& \cdots &{{Y_{1P}}\left( f \right)}&{{Y_{21}}\left( f \right)}& \cdots &{{Y_{M1}}\left( f \right)}
%\end{array}} \right]^{\rm T}}$, ${\bf{J}} = {\bf{G}}\left( {\left[ {1:P\;P + 1:\left( {M - 1} \right)P + 1} \right],\left[ {{l_1}, \cdots ,{l_K}} \right]} \right)$, we have the following equation according to (\ref{Yf}):
%\begin{align}\label{ABCS}
%{\bf{P}}\left( f \right) = {\bf{J}}\overline {\bf{S}} \left( f \right)
%\end{align}

\subsubsection{ Spatial Phase and Frequency matching algorithm}

%With sub-nyquist sampling, the received signal ${\overline Y _{mp}}\left( f \right)$ frequency aliasing will occur. The frequency estimates $\widehat{f} $ and $\overline{f}$ and  are similar. So, we can match the folding frequency and real frequency based the following rule
%\begin{align}\label{MatchFre}
%{f_i} = \left( {{\Omega _j} - 1} \right)\frac{{{f_s}}}{L} + \widehat {{f}}_i , when \widehat {{f}}_i  =  \overline{{f}}_j
%\end{align}
%
%Further, we obtain $\theta$ according to (\ref{Phi}).

We calculate the cross-correlation function of signal estimates ${{\bf{Z}}\left( f \right)}$ and ${\left( {{{\bf{X}}_1}} \right)_\Omega }\left( f \right)$. The absolute value of the cross-correlation matrix element has the following expression

%\begin{align}\label{Rxx}
%{{{R}}_{pm}} &= E\left\{ {{\bf{Z}}\left( f \right)\left( {{{\bf{X}}_1}} \right)_\Omega ^{\rm H}\left( f \right)} \right\} \\\nonumber
% &= \left[ {\begin{array}{*{20}{c}}
%{\sum\limits_{l = 1}^L {{{{B}}_{1l}}{S_{1l}}\left( f \right)} }\\
% \vdots \\
%{\sum\limits_{l = 1}^L {{{{B}}_{1l}}{S_{Kl}}\left( f \right)} }
%\end{array}} \right]{\left[ {\begin{array}{*{20}{c}}
%{\sum\limits_{k = 1}^K {{{{A}}_{1k}}{S_{k{\Omega _1}}}\left( f \right)} }\\
% \vdots \\
%{\sum\limits_{k = 1}^K {{{{A}}_{1k}}{S_{k{\Omega _c}}}\left( f \right)} }
%\end{array}} \right]^{\rm H}}
%\end{align}
%If $c = K$, ${{\bf{R}}}$ is a permutation matrix. So, we can match the folding frequency and real frequency based the following rule
%\begin{align}\label{MatchCor}
%{f_i} = \left( {{\Omega _j} - 1} \right)\frac{{{f_s}}}{L} + \widehat {{f_i}} ,when\;\left| {{{{R}}_{ij}}} \right| > 0.
%\end{align}
\begin{align}\label{Rpm}
\left| {{R_{ij}}} \right|
&= \left| {E\left\{ {{B_{1{l_i}}}{S_{i{l_i}}}\left( f \right){{\left( {\sum\limits_{k = 1}^K {{A_{1k}}{S_{k{\Omega _j}}}\left( f \right)} } \right)}^{\rm H}}} \right\}} \right|\\\label{mid}
&= \left| {E\left\{ {{S_{k{l_k}}}\left( f \right){{\left( {{S_{k{\Omega _j}}}\left( f \right)} \right)}^{\rm H}}} \right\}} \right|\\
&= \left\{ {\begin{array}{*{20}{c}}
{ > 0,when\;{l_i} = {\Omega _j}}\\
{ = 0,when\;{l_i} \ne {\Omega _j}}
\end{array}} \right.,1 \le i \le K,1 \le j \le c.
\end{align}
The conditions for the establishment of (\ref{mid}) have the signals are uncorrelated, the magnitudes of both ${B_{1{l_i}}}$ and ${A_{1k}}$ are 1.
If any of the two signal frequencies are in different frequency bands, we have $c = K$, or $c < K$. According to (\ref{Rpm}), we know that there is one absolute value of element is dominant in each row of ${\bf{R}}$. Further, the support index $\mathcal{S}$ of $\bf{H}$ is determined as following:
\begin{align}\label{Supp}
\mathcal{S}_i = \left( {i - 1} \right)L + {\Omega_j}, j = \mathop {\arg \max }\limits_j \left| {{R_{ij}}} \right|,1\leq i \leq K.
\end{align}
With known the support index $\mathcal{S}$, we obtain
\begin{align}\label{CS}
{\bf{W}}\left( f \right){\rm{ = }}{\bf{H}}\overline {\bf{S}} \left( f \right) = {{\bf{H}}_\mathcal{S} }{\overline {\bf{S}} _\mathcal{S} }\left( f \right) + {\bf{J}}{{\bf{I}}_{\bf{B}}}\widehat {\bf{N}}\left( f \right), f \in\mathcal{F}.
\end{align}
Then  we have the  least square solution of ${\overline {\bf{S}} _\mathcal{S} }\left( f \right)$
\begin{align}\label{MLES}
\overline{\bf{S}}_\mathcal{S}\left( f \right){\rm{ = }}{\bf{H}}_\mathcal{S} ^\dag {\bf{W}}\left( f \right).
\end{align}
We can gain the received signal's frequency ${\overline f _k}$ through $\overline{\bf{S}}_\mathcal{S}\left( f \right)$. Besides, there is a relationship between ${\overline f _k}$ and the original signal's frequency ${ f _k}$,
\begin{align}\label{MatchFreDir}
{f_k} = \left( {{{{\cal S}_k}\% L} - 1} \right)\frac{{{f_N}}}{L} + \overline {{f}}_k.
\end{align}

We can calculate $\theta_k$ through (\ref{Phi}). We outline the main steps of this individual estimate method for partial channels named algorithm JDFPI in table \ref{Alg4}.
\begin{table}[!t]
    \renewcommand{\arraystretch}{1.0}
    \caption{\textbf{Algorithm JDFPI}}\label{Alg4}
    \label{table_example}
    \centering
    \begin{tabularx}{8.4cm}{lX}
        \toprule
         1)&According to (\ref{ASub}),obtain $\phi$ applying the MUSIC, ESPRIT algorithm, and so on;\\
         2)&Compute ${ {{ {\bf{Z}} }}  }\left( f \right)$ according to  (\ref{ALS});\\
         3)&Apply the CTF algorithm to solve (\ref{BSub}) to obtain $\Omega $;\\
         4)&Compute ${\left( {{\overline {\bf{X}} _1}} \right)_\Omega }\left( f \right)$ according to  (\ref{BLS});\\
         5)&Determine  the support index $\mathcal{S}$ according to (\ref{Supp}); \\
         6)&Compute  ${\overline {\bf{S}} _\mathcal{S} }\left( f \right)$  according to (\ref{CS}); \\
         7)&Determine ${\overline f _k}$ through $\overline{\bf{S}}_\mathcal{S}\left( f \right)$ applying the MUSIC, ESPRIT algorithm, and so on; \\
         8)&Acquire ${ f _k}$ according to (\ref{MatchFreDir}); \\
         9)&Calculate  $\theta_k$ through (\ref{Phi});\\
        \bottomrule
    \end{tabularx}
\end{table}

\subsection{Algorithm Based on subspace decomposition}
If we calculate  the covariance matrix of ${\bf{W}}\left( f \right)$, and take advantage of the  subspace decomposition theory as \cite{Schmidt1986}, we will have similar conclusion:
\begin{align}\label{Orth}
{{\bf{a}}_l}\left( \phi  \right) \bot {{\bf{U}}_N},
\end{align}
where ${{\bf{U}}_N}$ is the noise subspace. The difference is ${{\bf{a}}_l}\left( \phi  \right) = {\bf{J}}\left( {{\bf{a}}\left( \phi  \right) \otimes {{\bf{B}}_l}} \right)$. Similarly, we can execute the steps of Algorithm JDFSD \cite{Liu2016}. It is worth pointing out that ${\bf{G}}$ and ${\bf{Y}}\left( f \right)$ in \cite{Liu2016} need to be replaced by  ${\bf{H}}$ and ${\bf{W}}\left( f \right)$, respectively. We name this method as algorithm based on subspace decomposition for partial channels (JDFSDPJ).
\subsection{Performance Analysis: Cram\'{e}r\text{-}Rao Bound}
Comparing the model (\ref{CS}) and model (11) in \cite{Liu2016} and noticing that (\ref{Noise}) holds, and making use of the conclusion of Section V equation (29) in \cite{Liu2016}, we have
\begin{align}\label{CRB4Sim}
{\rm{CRB}}_{sub}(sim) &= \frac{\sigma ^2}{{2T/L}}{\left( {\Re \left( {\left( {{{\bf{E}}^{\rm{H}}}{{\bf{P}}_{{{\bf{H}}_{\mathcal{S}}}}}{\bf{E}}} \right) \odot {\bf{R}}_{\overline {\bf{S}} }^{\rm{H}}} \right)} \right)^{ - 1}}\nonumber \\
&= \frac{{{\sigma ^2}}}{{2T}}{\left( {\Re \left( {\left( {{{\bf{E}}^{\rm{H}}}{{\bf{P}}_{{{\bf{H}}_{\mathcal{S}}}}}{\bf{E}}} \right) \odot {\bf{R}_{\bf{S}}^{\rm{H}}}} \right)} \right)^{ - 1}},
\end{align}
where ${{\bf{P}}_{{{\bf{H}}_{\mathcal{S}}}}} = {\bf{I}} - {{\bf{H}}_{\mathcal{S}}}{\bf{H}}_{\mathcal{S}}^\dag $, where ${\bf{H}}_{\mathcal{S}}^\dag  = {\left( {{\bf{H}}_{\mathcal{S}}^{\rm{H}}{{\bf{H}}_{\mathcal{S}}}} \right)^{ - 1}}{\bf{H}}_{\mathcal{S}}^{\rm{H}}$, ${\bf{E}} = \left[ {{{\bf{E}}_1}, \cdots ,{{\bf{E}}_K}} \right]$, ${{\bf{E}}_i} = \frac{{d{{{\bf{H}}_{{\mathcal{S}}_i}}}}}{{d{\phi _i}}}$.

\section{Simulation}
In this section, we present the numerical simulation results to illustrate the performance of the proposed algorithms. For the sake of comparison, we take JDFSD in \cite{Liu2016} as a representative of full structure, as JDFSD and JDFTD have the same performance. We set the receiver structure  as \cite{Liu2016}, and we take the all branches of the 1st sensor and the 1st branch of all sensors as our simplified structure. For the same reason mentioned in \cite{Liu2016},  we will only give the phase estimation simulation result rather than the DOA estimation simulation result in those simulations.

\subsection{Performance with noise}
Firstly, we will show our model can be solved by the proposed algorithm in different noise levels. In this subsection, the simulation scenario is the same as section VI-A in \cite{Liu2016}.

\begin{figure}[!t]
\centering
\includegraphics[width=2.5in]{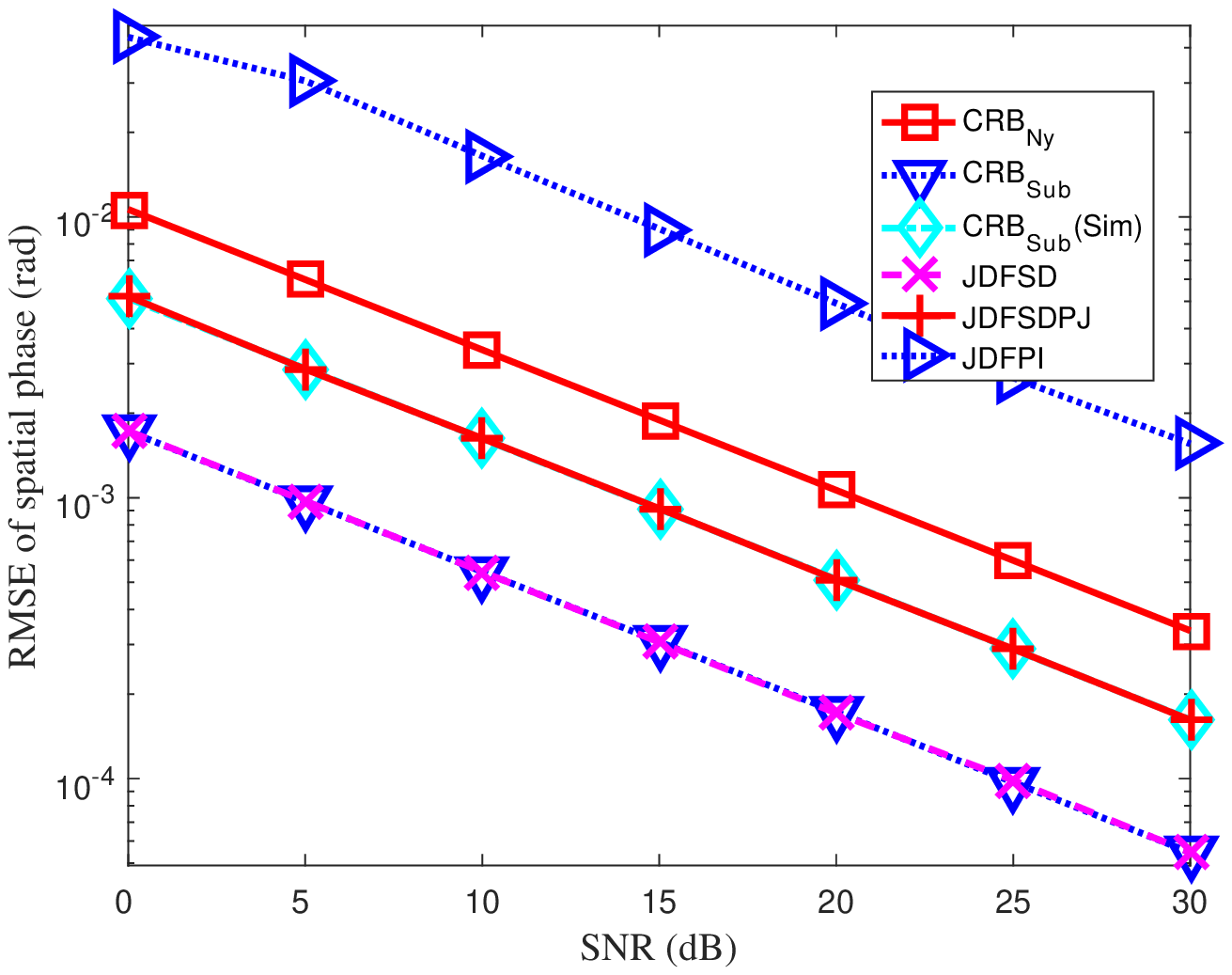}
\caption{RMSE of phase estimates versus SNR.}
\label{figPS}
\end{figure}

\begin{figure}[!t]
\centering
\includegraphics[width=2.5in]{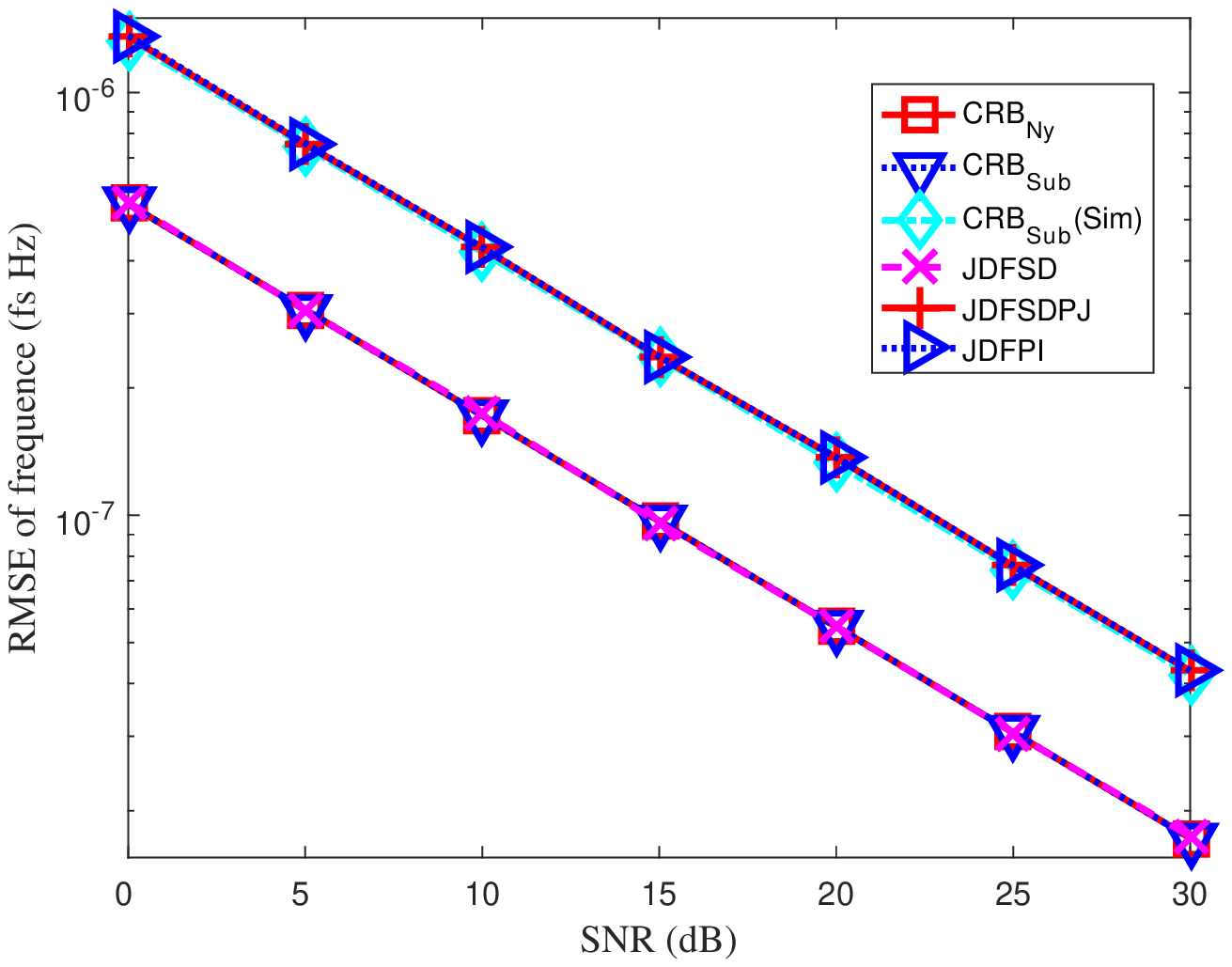}
\caption{RMSE of frequency estimates versus SNR.}
\label{figFS}
\end{figure}

Fig.\ref{figPS}-Fig.\ref{figFS} depict the RMSE versus  SNR in terms of spatial phase and frequency estimation, respectively.  Fig.\ref{figPS} shows that the phase estimation performance of algorithms JDFSDPJ and JDFPI improves with SNR, where JDFSDPJ achieves the $\textrm{CRB}_{sub}(Sim)$. The phase estimation performance of JDFSDPJ is better than that of JDFPI is because of jointly using the information in frequency domain and spatial domain. And we observe that $\textrm{CRB}_{sub}(Sim)$ lies between $\textrm{CRB}_{sub}$ and $\textrm{CRB}_{Ny}$. $\textrm{CRB}_{sub}(Sim)$ is higher than $\textrm{CRB}_{sub}$ is obvious.  The simplified structure use the jointly information from frequency domain and spatial domain. It leads to  a big improvement although the simplified structure have much less samplings comparing with Nyquist sampling. In Fig.\ref{figFS} demonstrates that the frequency estimation performances of JDFSDPJ and JDFPI can achieve the $\textrm{CRB}_{sub}(Sim)$, which is certainly  higher than $\textrm{CRB}_{sub}(Sim)$ because of using less branches.

%\begin{figure}[!t]
%\centering
%\includegraphics[width=2.5in]{DOAVsSNR.eps}
%\caption{RMSE of DOA estimates versus SNR.}
%\label{figDS}
%\end{figure}

\subsection{Performance with various signal number}
\begin{figure}[!t]
\centering
\includegraphics[width=2.5in]{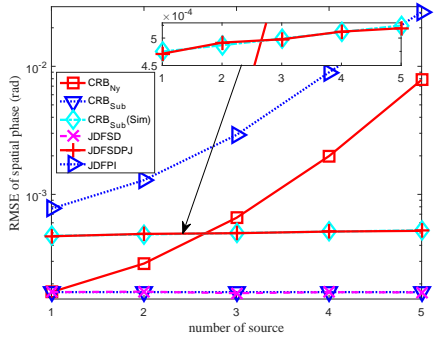}
\caption{RMSE of phase estimates versus number of source.}
\label{figPK}
\end{figure}

\begin{figure}[!t]
\centering
\includegraphics[width=2.5in]{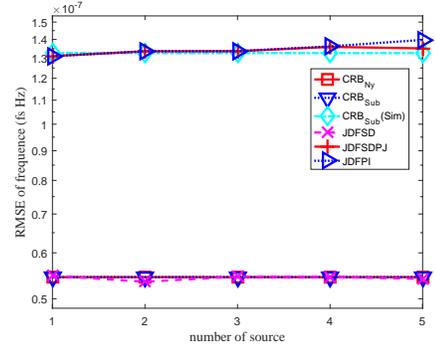}
\caption{RMSE of frequency estimates versus number of source.}
\label{figFK}
\end{figure}

In this  subsection, we will investigate the estimation performance when the signal number changes as section VI-C in \cite{Liu2016}.
Fig.\ref{figPK} shows that the phase (DOA) estimation performance of algorithm JDFSDPJ is slightly influenced by the signal number and achieves $\textrm{CRB}_{sub}(Sim)$, however JDFPI is influenced by the signal number. This is due to the former jointly using the information from frequency domain and spatial domain. To some degree, it maintains good robustness in terms of the number of signals as JDFSD. Without doubt the performance of JDFSDPJ is still worse than $\textrm{CRB}_{sub}$. Fig.\ref{figFK} shows that the frequency estimation performances of algorithms JDFSDPJ and JDFPI are not influenced by the signal number and can reach  $\textrm{CRB}_{sub}(Sim)$.

%\begin{figure}[th]
%\centering
%\includegraphics[width=80mm]{DOAVsK.eps}
%\caption{RMSE of DOA estimates versus  number of source.}
%\label{figDK}
%\end{figure}

\section{Conclusions}
In this paper, we designed an simplified array receiver architecture by introducing sub-Nyquist sampling technology. We realized the joint DOA and frequency estimation under lower sampling rate. Although the estimate precision of using partial channels is worse than that of using full channels, the former has lower equivalent sampling rate and hardware complexity. And increase time of sensing will enhance its estimation performance. The simulations demonstrated that the joint algorithm can closely match the CRB according to noise levels and source number as well.

%\section*{Acknowledgment}
%The authors would like to thank...

\ifCLASSOPTIONcaptionsoff
  \newpage
\fi

\bibliographystyle{IEEEtran}
\bibliography{IEEEabrv,ML}
%\bibliography{ML}

\end{document}